\documentclass[conference]{IEEEtran}
\IEEEoverridecommandlockouts
\usepackage{graphicx}
\usepackage{amsmath}
\usepackage{cite}
\usepackage{url}

\begin{document}

\title{EPC Framework for BESS Projects}

\author{\IEEEauthorblockN{Zeenat Hameed, Chresten Træholt}
\IEEEauthorblockA{Technical University of Denmark, Department of Wind and Energy Systems\\
Kongens Lyngby, Denmark\\
zeeha@dtu.dk, ctra@dtu.dk}}

\maketitle

\begin{abstract}
Battery Energy Storage Systems (BESS) are critical for modern power networks, supporting grid services such as frequency regulation, peak shaving, and black‐start. Delivering a BESS under an Engineering, Procurement, and Construction (EPC) model requires a concise methodology that balances regulatory compliance, technical details, and schedule efficiency. This paper presents a streamlined, five‐step EPC framework covering feasibility assessment, permitting, procurement, construction, and commissioning. A Danish demonstration (the BOSS project on Bornholm) serves as a case study.   
\end{abstract}


\section{Introduction}
Battery Energy Storage Systems (BESS) play a pivotal role in balancing variable renewable generation, providing ancillary services such as frequency containment reserve (FCR) and automated frequency restoration (aFRR), and offering energy arbitrage opportunities. However to provide these services, BESS must integrate with transmission or distribution networks, adhere to stringent safety and fire codes, and demonstrate performance through factory acceptance tests (FAT), site acceptance tests (SAT), and prequalification tests. 

Current academic and industry studies often focus on optimization algorithms, control strategies, and simulation models for BESS operation. While such work is invaluable, it does not address the real‐world challenges of project execution. BESS projects remain relatively new and complex to deploy; to scale up the number of installations, practitioners need clear processes and procedures. Typically, this implementation knowledge remains with a small number of specialized engineers and project managers. Moreover, BESS projects differ from large‐scale conventional generators and other infrastructure projects because they can be delivered by smaller commercial entities, cooperatives, or independent developers, in contrast to traditional coal‐fired or large gas‐fired plants, which require extensive corporate resources.

However, detailed understanding of what it takes to run a BESS project—including timelines, complexities, necessary steps, budgets, and stakeholder roles—remains ambiguous for many, even within the energy domain. While simulation studies and control algorithms are critical components, they must be viewed as part of the larger project lifecycle rather than isolated topics. To address these gaps, this paper focuses specifically on the Engineering, Procurement, and Construction (EPC) process for BESS projects, highlighting each phase and critical tasks.

Using Denmark as a case study, we detail the step‐by‐step EPC process and present a 1 MW/1 MWh BESS project in Bornholm as an illustrative example of how this methodology applies in practice. The paper is structured as follows: Section I provides the Introduction; Section II presents an overview of the EPC process for BESS projects; Section III applies this EPC process to the 1 MW/1 MWh BESS project in Bornholm. Section IV ends with conclusions.

\section{BESS-EPC Process Overview}
An EPC (Engineering, Procurement, and Construction) process defines the end‑to‑end sequence of activities required to deliver a BESS project from initial concept through ready‑for‑operation. It is important because it structures workflows, assigns responsibilities, and ensures that technical, commercial, and regulatory requirements are met on time and within budget. Without a clear EPC framework, projects risk delays, cost overruns, and misalignment among stakeholders.

In this paper, the EPC process encompasses five key steps:

\begin{enumerate}
\item {Feasibility Assessment}
\item {Permitting }
\item {Procurement}
\item {Construction}
\item {Commissioning}
\end{enumerate}

\subsection{Feasibility Assessment}
Feasibility studies are the foundation of any EPC project. They evaluate whether a BESS project would be a viable business venture in the specified geography. Key activities include:
\begin{itemize}
\item \textit{Business Case Evaluation}: Estimate capital expenditures (CAPEX), operational expenditures (OPEX), revenue streams, and return‑on‑investment (ROI). Financial models consider local market prices, incentives, and grid tariffs.
\item \textit{Site Selection}: Screen candidate sites based on proximity to electrical infrastructure, available land, zoning, environmental constraints, and land ownership. 
\item \textit{Preliminary Design}: Develop high‑level Single‑Line Diagrams (SLDs) showing Point of Connection (PoC) at the appropriate voltage level
\item \textit{High‑Level Cost Estimate}: Incorporate DSO / TSO‑provided grid reinforcement costs (e.g.cable upgrades, transformer replacements) and a civil works budget. Deliver a site selection report recommending the optimal location based on technical, financial, and schedule criteria.
\end{itemize}

\subsection{Permitting and Regulatory Approvals}
Permitting ensures legal authorization to proceed with construction and operation. Key steps include:
\begin{itemize}
\item \textit{Grid Connection Permits}: Submit a Request for Connection (RFC) to the local DSO or TSO, depending on voltage level.The DSO/TSO issues a binding connection offer detailing maximum power injection/withdrawal, reinforcement obligations, connection fees, and a timeline.
\item \textit{Power‑Quality Checks}: Finalize harmonic and flicker analysis using actual feeder impedances. If THD exceed thresholds, specify mitigation measures (tuned passive filters, active filters, or ramp‑rate limits).
\item \textit{Building Permit}: Provide site layout plans (container placement, transformer pad, fencing, access roads), elevation drawings (container height, transformer height), structural slab calculations, stormwater management, and noise assessment.
\item \textit{Fire Safety Permit}: Prepare fire‑safety documentation , fire suppression design, smoke/heat detection, fire‑load calculations, smoke extraction, and emergency response plan. 
\item \textit{Land and Easement Documentation}: Provide title deed or lease agreement. 
\end{itemize}

\subsection{Procurement and Supplier Management}
Once permits are secured, procurement begins:
\begin{itemize}
\item \textit{RFQ Preparation}: Develop a request for quotation (RfQ) outlining technical specifications, commercial terms, liquidated damages, and warranty requirements.
\item \textit{Supplier Prequalification}: Score vendors on technical compliance, commercial competitiveness, delivery timeline, warranty/service , and local support, and financial stability .
\item \textit{Supply Chain Management}: Track lead times for batteries, inverters, transformers, switchgear, and auxiliary equipment. Coordinate factory acceptance test (FAT) schedules at vendor facilities. Address any non‑conformances before shipment.
\item \textit{Logistics and Customs}: For imported equipment, handle customs clearance, and transport to port. Plan for special permits (e.g., overweight/oversize loads for transformers, crane capacities).
\end{itemize}

\subsection{Construction and Installation}
Construction brings the engineered design to life:
\begin{itemize}
\item \textit{Site Mobilization}: Establish site office, security fencing, portable facilities, and temporary lighting. 
\item \textit{Civil Works}: Excavate cable trenches. Implement erosion control. 
\item \textit{Electrical Installation}: Place transformers on epoxy‑coated bunds. Install MV switchgear. Place BESS containers on antivibration pads
\item \textit{Communications Infrastructure}: Install ethernet connection and develop communication set-up
\end{itemize}

\subsection{Testing, Commissioning, and Handover}
Commissioning validates system performance and readiness:
\begin{itemize}
\item \textit{Site Acceptance Test (SAT)}: Post‑installation, test protection relays. Validate inverter synchronization. Measure THD and flicker  under steady power  and full discharge conditions. Conduct full charge/discharge cycle to calculate round‑trip efficiency. Execute frequency support test. Test communications failover. Verify SCADA integration.
\item \textit{Prequalification}: Submit prequalificationn test results to the TSO for ancillary services. 
\item \textit{Market Integration}: Coordinate with a Balance Responsible Party (BRP) for trading access. Configure telemetry and control setpoints
\end{itemize}

\section{BESS-EPC framwwork implementation - A Danish Case-Study}

The Bornholm Smartgrid Secured (BOSS) project is used as a case-study in this paper. BOSS was a collaborative demonstration led by the Technical University of Denmark (DTU) that focused on the installation and operation of a 1 MW/1 MWh grid‐connected BESS on the Danish island of Bornholm. 

\subsection{Phase 1: Feasibility Assessment}

A detailed feasibility assessment was conducted to establish the business case, shortlist potential installation sites, and develop a connection schematic for the 1MW/1MWh BESS on Bornholm (DK2). Key sub‐steps are described below.

\subsubsection{Business Case Evaluation}
\begin{figure}[!ht]
  \centering
  \includegraphics[width=\linewidth]{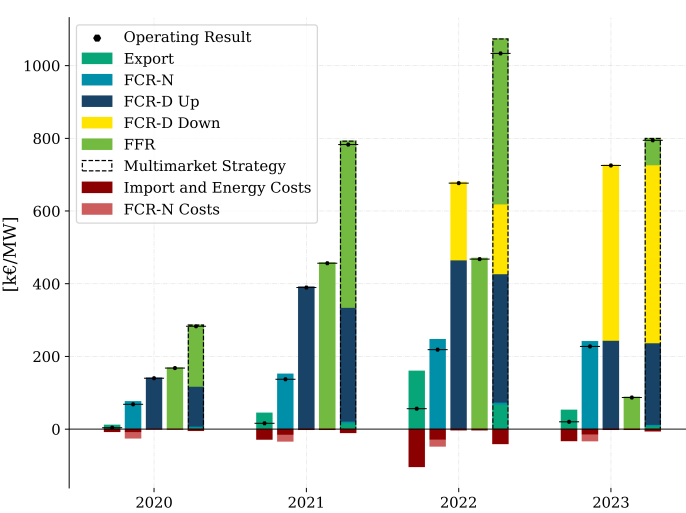}
  \caption{Annual operating result and revenue breakdown for a 1MW/1MWh BESS in DK2 (Bornholm)}
  \label{fig:business}
\end{figure}

A revenue model was developed to quantify the expected annual operating income of a 1MW/1MWh BESS in the DK2 bidding zone. Figure \ref{fig:business} shows the breakdown of operating results from primary grid services—FCR-N, FCR-D up/down, FFR—as well as export revenues and net import/energy costs, for the period 2020–2023. These results illustrate that, under the multi-market strategy, annual net revenues can exceed 800 kEUR/MW at peak market conditions in DK2.

\subsubsection{Site Selection}
\begin{figure}[!ht]
  \centering
  \includegraphics[width=\linewidth]{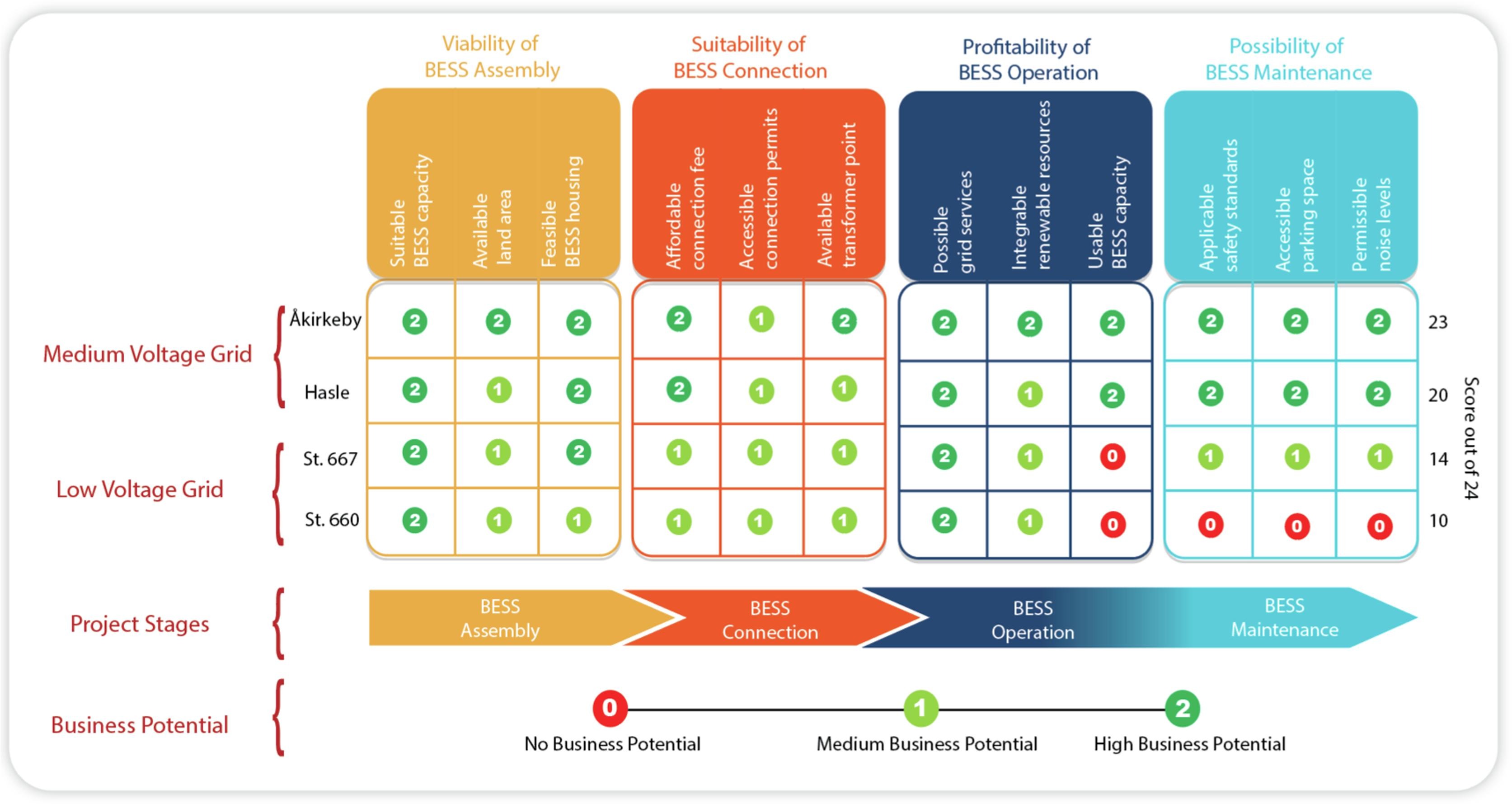}
  \caption{Site‐selection scoring matrix for Bornholm BESS candidates. Four sites (two on Medium‐Voltage grid, two on Low‐Voltage grid) were scored across four project‐stage pillars}
  \label{fig:site_selection}
\end{figure}

A systematic site‐selection framework was defined to evaluate four potential locations on Bornholm, using four criteria: (1) Viability of BESS Assembly (land availability, container footprint, soil conditions), (2) Suitability of BESS Connection (affordable connection fee, accessible permit process, available transformer), (3) Profitability of BESS Operation (proximity to network congestion points, renewable supply, local load profiles), and (4) Possibility of BESS Maintenance (safety standards, access roads, noise constraints). Each criterion was scored on a 0–2 scale (0 = no potential, 1 = medium, 2 = high), for a maximum possible score of 24. Figure \ref{fig:site_selection} summarizes the results for the four shortlisted sites: Aakirkeby, Hasle, St. 667, and St. 660.

As shown in Fig. \ref{fig:site_selection}, \textbf{Aakirkeby} scored 23/24, making it the preferred location. Key reasons included:
\begin{itemize}
  \item \textit{Land Availability and Zoning:} A municipally-owned brownfield parcel was immediately available with proper zoning for containerized installations.
  \item \textit{Grid Access:} Aakirkeby lies adjacent to a 60/10 kV substation that already had space reserved for BESS integration.  
  \item \textit{Operational Synergies:} High local wind output and occasional congestion on DK2 grid suggested strong ancillary service potential.  
  \item \textit{Maintenance:} Proximity to paved roads and minimal environmental/noise restrictions.  
\end{itemize}

\subsubsection{High‐Level Connection Schematic}
\begin{figure}[!ht]
  \centering
  \includegraphics[width=\linewidth]{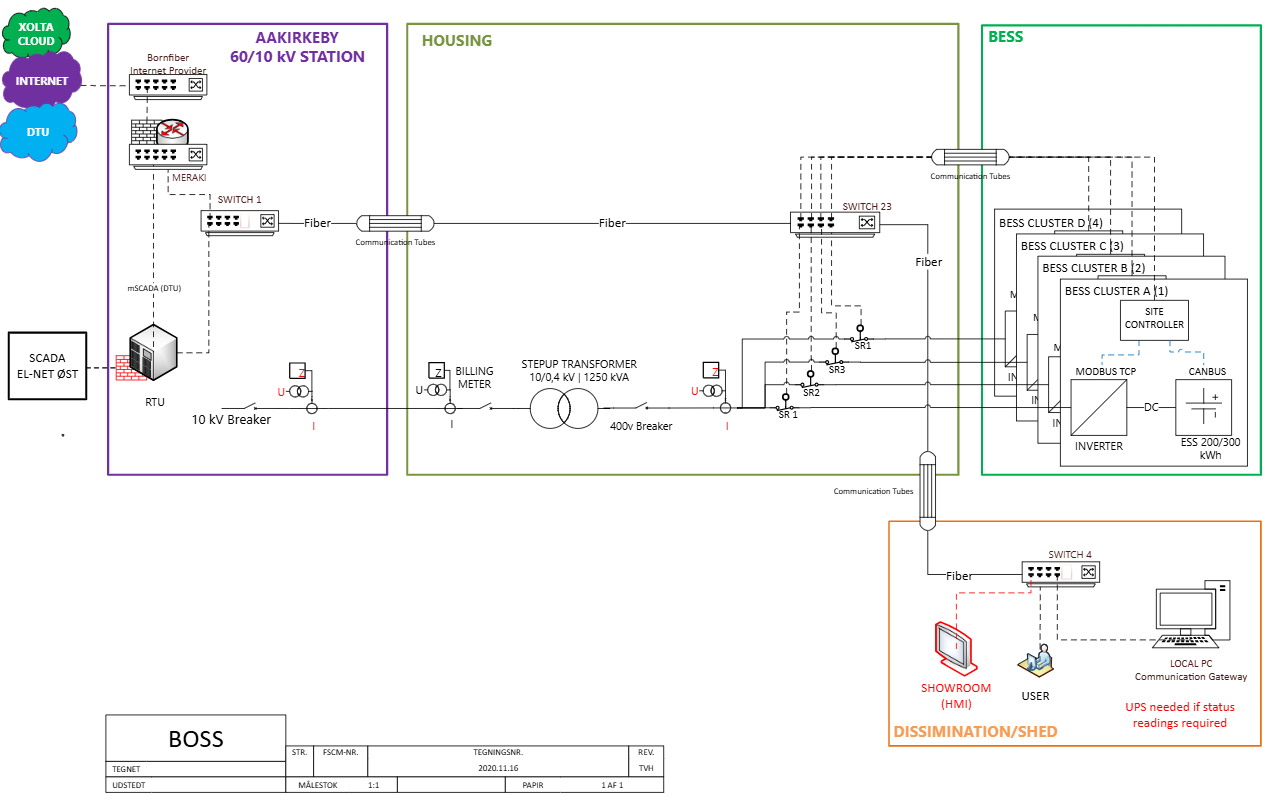}
  \caption{Preliminary single‐line diagram for Aakirkeby BESS}
  \label{fig:sld}
\end{figure}

Once Aakirkeby was chosen, a preliminary single‐line diagram (SLD) was drafted to capture the electrical configuration and communications architecture (see Fig. \ref{fig:sld}). The SLD shows:
\begin{itemize}
  \item A \textbf{60/10 kV substation} on the left (purple outline) with DA‐grade RTU and SCADA link to DTU PowerLabDK and BESS supplier Cloud.  
  \item A new \textbf{10 kV breaker} feeding through a step‐up/step‐down 10/0.4 kV transformer (1250 kVA), with on‐load metering for billing.  
  \item Medium‐Voltage switchgear (SR1, SR2, SR3) interfacing the 10 kV bus to four parallel BESS clusters (A–D), each consisting of a 200 kW/300 kWh inverter + DC battery module. Within each cluster, a local “Site Controller” manages Modbus‐TCP to inverter and CANbus to battery.  
  \item A dedicated \textbf{dissemination shed} (orange outline) containing an HMI “Showroom” (for live demonstrations) and a local PC for status readouts; this block is cabled back to “Switch 4” at the substation via fiber.
\end{itemize}

In Fig. \ref{fig:sld}, the dashed communication tubes indicate fiber‐optic links from the RTU at the 60/10 kV station to the Site Controller in each BESS cluster, ensuring redundant telemetry for both operational control and research data collection.

\subsubsection{High‐Level Connection Schematic}
After selecting Aakirkeby as the preferred site, the project team evaluated four possible connection categories—A‐high, A‐low, B‐high, B‐low, and C—each corresponding to a different voltage level and physical connection point in the local grid. 

\begin{figure}[!ht]
  \centering
  \includegraphics[width=\linewidth]{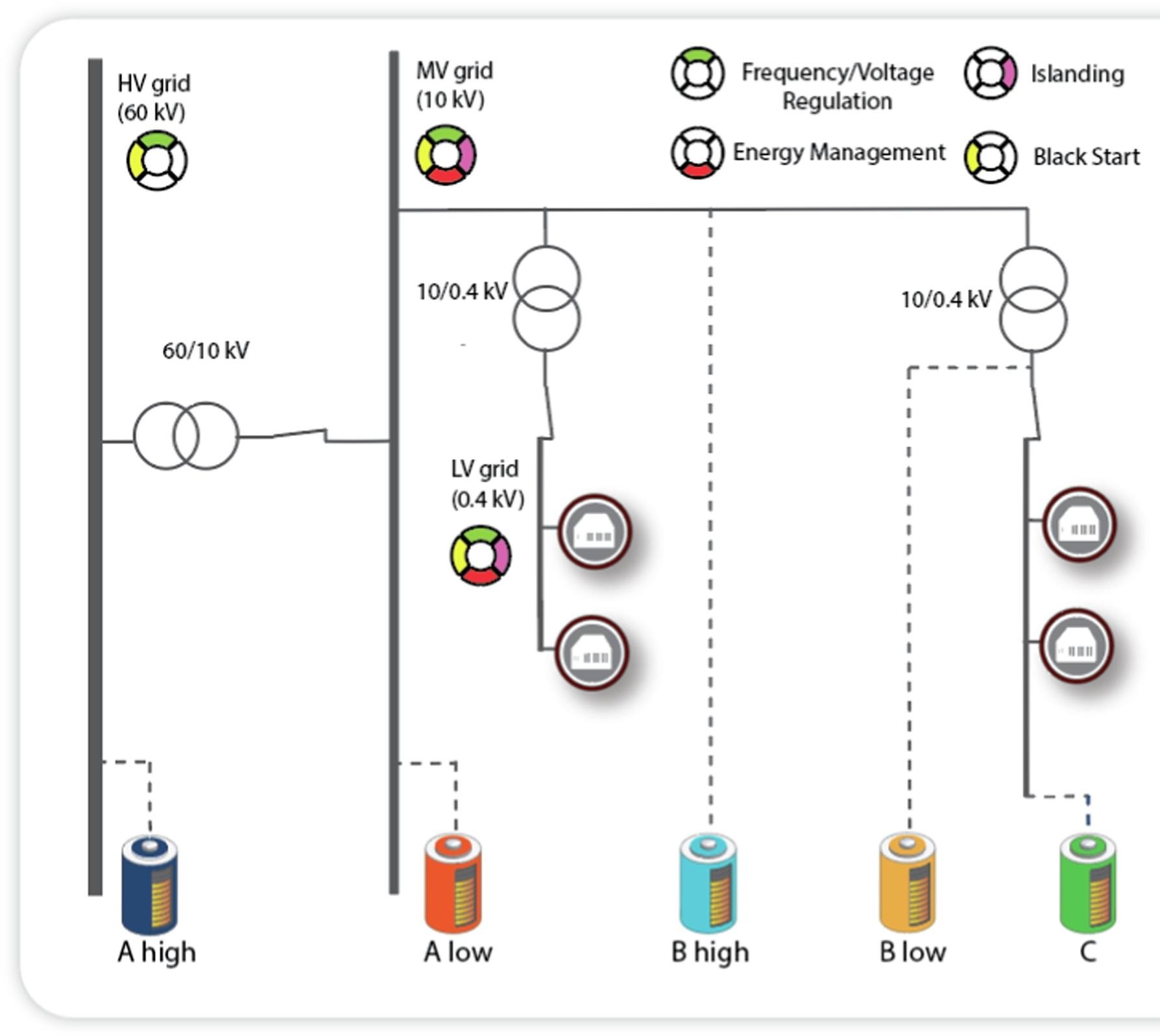}
  \caption{BESS connection categories in Bornholm’s network.}
  \label{fig:connections}
\end{figure}

Because the BESS demonstration aimed to showcase low‐voltage islanding and black‐start, the Aakirkeby site ultimately connected under \textbf{B‐high} (10 kV at a 10 kV station) for its balance of cost and operational flexibility. However, the project team modeled all five categories when compiling the high‐level cost estimate below.

\subsubsection{High‐Level Cost Estimate for Grid Connection}
Using RAH Net’s standard “Tilslutningsbidrag” (connection fee) rates (effective January 1 2024), the fees for each category are summarized in Table \ref{tab:connection_fees}. These fees include VAT and apply to new installations or significant upratings:

\begin{table}[!ht]
  \centering
  \footnotesize
  \caption{Grid Connection Fees by Category (RAH Net, Jan 1 2024).}
  \label{tab:connection_fees}
  \begin{tabular}{@{}l p{5cm} r@{}}
    \hline
    \textbf{Category}                   & \textbf{Fee (DKK)} \\ 
    \hline
    A-high (30–60 kV)                          & 700 000 DKK/MVA   \\
    A-low (10 kV of 60 kV)              & 1 280 000 DKK/MVA \\ 
    B-high (10 kV)                         & 850 000 DKK/MVA   \\ 
    B-low (0.4 kV at 10 kV)          & 1 310 DKK/A       \\
    C (0.4 kV cable cabinet)              & 1 360 DKK/A       \\ 
    \hline
  \end{tabular}
\end{table}

For a 1 MW BESS (requiring a 1.25 MVA transformer for MV options or approximately 1 804 A at 0.4 kV for LV options), the approximate connection fees are shown in the table. Although \emph{A‐high} (875 kDKK) is the least expensive, and \emph{A‐low} (1 600 kDKK) is the next‐best option, the BOSS team chose \textbf{B‐high (1 062 kDKK)} at the 10 kV station. This allowed demonstration of both MV grid services and 0.4 kV islanding (via an on‐site 10/0.4 kV transformer), balancing cost and operational flexibility.

\bigskip

\subsection{Permitting and Regulatory Approvals}

\begin{figure}[!ht]
  \centering
  \includegraphics[width=0.85\linewidth]{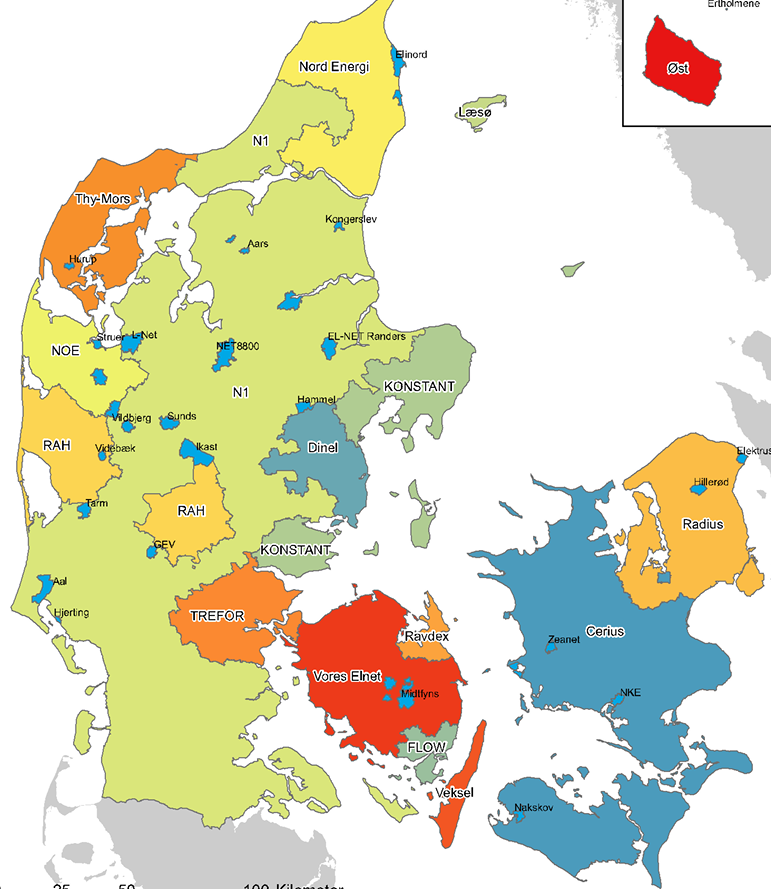}
  \caption{DSO service territories in Denmark}
  \label{fig:dso_map}
\end{figure}

Once the feasibility assessment confirmed Aakirkeby as the preferred site, the permitting phase began in earnest. The first step was to secure a formal \emph{grid connection permit} with the appropriate Distribution System Operator (DSO). In Denmark, the map in Figure \ref{fig:dso_map} shows how each region is served by a different DSO - each responsible for handling connection requests within its territory. Because Bornholm falls under the \emph{Trefor Elnet} network (labeled “Øst”), the BOSS project team submitted their Request for Connection (RFC) directly to Trefor Elnet’s local office.

In the RFC packet, the following items were submitted to Trefor Elnet:
\begin{itemize}
  \item The single‐line diagram (SLD) in its final engineering detail, showing the point of connection at 10 kV RMU (ring main unit)
  \item A completed copy of Annex B1.1 (under “Documentation for Energy Storage Facilities Type A”) from the Danish technical regulation, containing inverter specifications, storage medium details, and initial protection settings (over-/undervoltage, over-/underfrequency, and power‐quality parameters). 
\end{itemize}

\paragraph{Power‐Quality Documentation}
As mandated by Annex B1.1–B1.2, the BESS must demonstrate compliance with harmonic, flicker, and DC‐injection limits at the point of common coupling (PCC). The key power‐quality checks submitted were related to DC Content, Current Unbalance, Rapid Voltage Changes, Flicker Contribution, and Harmonic Overtone Limits.

Trefor Elnet reviewed and accepted these results. The processing time for grid connection was upto a year.

\paragraph{Building and Fire Safety Permits}
While the grid connection application was under review, the the \emph{building permit} package for the local municipality was submitted. Submissions included:
\begin{itemize}
  \item A high‐resolution site plan (Figure \ref{fig:site map}), overlaid on aerial imagery, showing the precise location of each battery container.   
  \item A noise impact report, confirming that combined transformer and inverter fan noise at the property line remains below 45 dB(A) during steady discharge.  
  \item Stormwater management plan, showing site grading draining away from structures and connecting into nearby municipal drainage lines.  
\end{itemize}

\paragraph{Municipal Review and Permit Issuance}
Between 4 - 6 months of the permitting phase, the municipality reviewed the building application and the fire department reviewed the fire‐safety package. On Bornholm, local regulations require a joint Site Inspection to confirm that slab elevations, drainage gradients, and noise mitigation measures align with the approved permit. At the conclusion of this review, both the building permit and fire‐safety certificate were issued.

\begin{figure}[!ht]
  \centering
  \includegraphics[width=\linewidth]{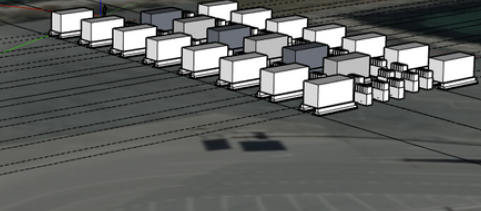}
  \caption{3D site model (SketchUp screenshot)}
  \label{fig:site_map}
\end{figure}
\subsection{Procurement and Supplier Management}

Because BOSS was conceived as a research‐driven demonstration project, the BESS supplier selection occurred during the initial consortium‐formation stage rather than through a standalone competitive bid. As a result, the formal RFQ issued by the project team focused on refining technical interfaces and securing firm delivery schedules, rather than inviting a broad array of new bidders.

\subsection{Construction and Installation}

Construction of the BOSS BESS was executed jointly by the contracted BESS integrator (battery and inverter supplier) and DTU PowerLab DK’s field engineers. 

\subsection{Testing, Commissioning, and Handover}

\begin{figure}[!ht]
  \centering
  \includegraphics[width=0.8\linewidth]{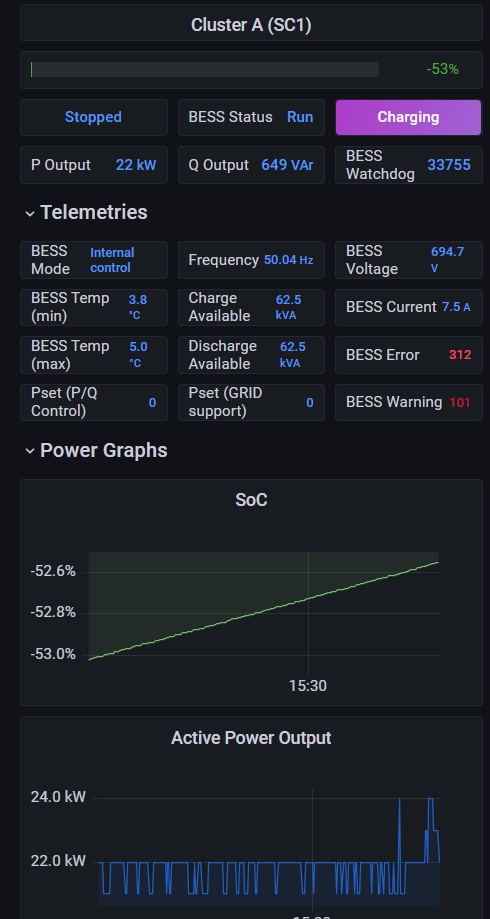}
  \caption{PowerLab DK’s real-time BESS control dashboard.}
  \label{fig:status_dashboard}
\end{figure}

Commissioning was led by DTU PowerLab DK, leveraging a custom control system developed in house. This proprietary controller, deployed on an industrial PC inside the AC control building, orchestrates all BESS clusters via Modbus TCP/IP and CANbus communications, executes grid-support algorithms, and interfaces with the DSO RTU. Figure \ref{fig:status_dashboard} shows a snapshot of the live control dashboard used by PowerLab DK engineers to monitor cluster states and power flows.

For SAT, the PowerLab DK control system was used to issue remote commands to each cluster. Three consecutive charge/discharge cycles at a C/2 rate measured round-trip efficiency. BMS telemetry was validated—ensuring cell‐voltage spread, temperature sensor accuracy within ±2 °C, and communication failover (fiber redundancy) in under 200 ms.

Figure \ref{fig:status_dashboard} illustrates how PowerLab DK engineers observe real-time SoC trends (Cluster A at –53 percent indicates charging).

\section{Conclusion}
This paper has outlined a concise EPC framework for BESS projects, structured into five sequential phases. The BOSS case demonstrates that with a 10 kV (Bhj) connection - chosen to highlight medium-voltage islanding and black start - the project achieved on-budget construction and on-time commissioning by adhering to the outlined EPC framework. Early DSO collaboration and meticulous power‐quality compliance were particularly critical in mitigating technical and schedule risks. The consortium‐driven procurement model reduced administrative overhead, while the in‐house control solution from PowerLab DK enabled seamless transition from testing to commercial operation. Future work will evaluate long‐term operational data (e.g., degradation rates, actual revenue streams) and expand the framework to larger multi‐site implementations across different regulatory environments.  

\section*{Acknowledgment}
The authors are thankful to PowerLabDK for the collaboration and support.

\end{document}